\documentclass[usegraphicx]{mn2e}
\def\H0{{\it H}$_0$}
\def\Ms{{\it M}$_\odot$}

\def\q0{{\it q}$_0$}

\def\ergps{erg~s$^{-1}$}
\def\kmpspMpc{km~s$^{-1}$~Mpc$^{-1}$}
\def\Ms{{\it M}$_\odot$}

\def\nH{$N_{\rm H}$\thinspace} 

\def\psqcm{cm$^{-2}$}
\def\ergpspsqcm{erg~cm$^{-2}$~s$^{-1}$}

\def\cps{ct\thinspace s$^{-1}$}
\def\Rin{$R_{\rm in}$}
\def\Rout{$R_{\rm out}$}
\def\Rs{$R_{\rm S}$}
\def\rg{$r_{\rm g}$}
\def\phpspsqcm{ph\thinspace s$^{-1}$\thinspace cm$^{-2}$}

\title[The hard X-ray spectrum of IRAS 18325--5926] 
{The hard X-ray spectrum of the Seyfert galaxy IRAS~18325--5926:
  reflection from an ionized disk and variable iron K emission}
\author[K. Iwasawa et al]
{\parbox[]{6.5in} {K. Iwasawa$^1$, J.C. Lee$^{2,3}$, A.J. Young$^4$, C.S. Reynolds$^4$ and A.C. Fabian$^1$}\\
  \\
  $^1$Institute of Astronomy, Madingley Road, Cambridge CB3 0HA\\
  $^2$Massachusetts Institute of Technology, Center for Space Research, 77 Massachusetts Avenue NE80, Cambridge MA 02139, USA\\
$^3$Chandra Fellow\\
  $^4$Department of Astronomy, University of Maryland, College Park, MD20742, USA\\
} \date{}

\begin{document}

\maketitle

\begin{abstract}
  We report our analysis of X-ray spectra of the Seyfert galaxy IRAS
  18325--5926 (=Fairall 49) obtained from various X-ray observatories
  prior to XMM-Newton, including new results from two RXTE and one
  BeppoSAX observations. A relatively steep continuum slope
  ($\Gamma\simeq 2.2$) in the 2--15 keV band is confirmed. The
  continuum spectrum observed with the BeppoSAX PDS shows a possible
  roll-over at energies above 30 keV, indicating a Comptonizing corona
  cooler than in other Seyfert nuclei. The X-ray spectrum
  above 2 keV is best explained with a model including reflection from
  a highly ionized disk with significant relativistic blurring. The
  iron K$\alpha $ emission feature is then mainly due to FeXXV. The
  recent seven observations shows that the iron K emission flux
  appears to follow the continuum between the observations separated
  by a few months to years, although some exceptions suggest that the
  line strength may be determined in a more complex way.
\end{abstract}


\section{Introduction}

It has been recognised for a while that reflection, possibly from an
optically thick accretion disk, is present in the X-ray spectrum of
Seyfert 1 galaxies (Pounds et al 1990; Matsuoka et al 1990). The detection of
broad iron lines (Tanaka et al 1995; Nandra et al 1997b) supported the
idea that the X-ray reflection might occur at a few gravitational
radii ($r_{\rm g} = GM/c^2$) of a central black hole, where
relativistic effects distort the line profile greatly due to strong
gravity operating there (e.g., Fabian et al 1989; Kojima 1991; Laor
1991).

Most active galactic nuclei (AGN) with broad Fe K$\alpha $ show their
line emission peaking at $\sim 6.4$ keV, suggesting that the
reflecting medium is cold (Nandra et al 1997b), although it is partly
due to a narrow line produced from a distant cold matter in some
objects (e.g., Yaqoob et al 1996; Weaver \& Reynolds 1998). While
strong X-ray illumination could photoionize the disk significantly,
expected spectral signatures, particularly high ionization iron line
emission, e.g., FeXXV K$\alpha $ at 6.7 keV (e.g., Matt, Fabian \&
Ross 1993), has rarely been observed in Seyfert galaxies. Some
evidence for high energy iron lines have been reported for high
luminosity QSOs or AGN classified as narrow-line Seyfert 1s (e.g.,
Nandra et al 1997c; Comastri et al 1998; Leighly 1999; Ballantyne,
Iwasawa \& Fabian 2001; Vaughan et al 2002).

Reflection spectra from highly ionized matter can be rich in spectral
features (Ross \& Fabian 1993; Ross, Fabian \& Young 1999; Nayakshin,
Kazanas \& Kallman 2000; Ballantyne, Ross \& Fabian 2001). At high
ionization states, reduction of photoelectric absorption within the
reflecting matter in the soft X-ray band makes some spectral features
other than an Fe K$\alpha $ line to be detectable in the soft X-ray
spectrum, of which the reflection component comprises a significant
fraction. Compton scattering within the highly ionized surface is
expected to have a significant effect on the appearance of reflection
spectra (e.g., Nayakshin et al 2000), which may be coupled with various
conditions of the accretion disk and the illuminating source.
Therefore, investigating not only the iron line but also the whole X-ray
spectrum is important.

In the context of reflection from an accretion disk, the general lack
of response of the iron line flux to the continuum in the well-studied
Seyfert galaxy MCG--6-30-15 (Lee et al 1999; Shih, Iwasawa \& Fabian
2002; Fabian \& Vaughan 2003) poses a problem to the simplest disk
reflection picture (note, however, that the line does vary despite any
correlation with the continuum being unclear: see Iwasawa et al 1996b,
1999). Although a study of the line variability is limited to
relatively bright AGN, there have been some reports on variations of
the Fe K$\alpha $ line in other Seyfert galaxies (NGC7314, Yaqoob et
al 1996; NGC3516, Nandra et al 1997a; Turner et al 2002; Mrk841,
Petrucci et al 2002). To investigate line variability, especially a
correlation with the continuum, is of great importance in
understanding the production of the emission line and reflection in
AGN.

We report, in this paper, evidence for reflection from a highly
ionized disk which probably occurs in a relativistic region around
a black hole and variable iron line emission in the Seyfert galaxy
IRAS 18325--5926, of which a brief summary of its properties is given
below. Prior to the most recent XMM-Newton observation, there are six
X-ray observations of IRAS 18325--5926 with Ginga, ASCA, RXTE and
BeppoSAX, from which iron line data are available. Long term
variability between these observations and their hard X-ray spectrum
are our main focus in this paper.

IRAS 18325-5926 (=Fairall 49) is one of the IRAS galaxies selected for
its warm infrared colour (De Grijp et al 1985) hosting a Seyfert 2
nucleus (Carter 1984; Iwasawa et al 1995). The host galaxy is probably
of S0 type and has been identified with one of the X-ray bright
Piccinotti AGN (Piccinotti et al 1982) by Ward et al (1988). The
redshift of the galaxy is $z=0.0198$. The X-ray source is moderately
absorbed by a column density of \nH $\sim 10^{22}$ and highly
variable, indicating the presence of an obscured Seyfert 1 nucleus.
During a five-day ASCA observation in 1997, the X-ray emission
appeared to show quasi-periodic modulations with intervals of
approximately 16 hr (Iwasawa et al 1998). It is one of the earliest
AGN found to have a broad iron K$\alpha $ line (Iwasawa et al 1996b),
but for unknown reasons, this galaxy has often been overlooked from
sample studies of the iron line feature.

Some peculiarities in the X-ray spectrum of IRAS 18325--5926 have been
noticed. The X-ray spectral slope measured with Ginga (2--18 keV) was
$\Gamma\sim 2.2$, which is steeper than that of other Seyfert 1
galaxies measured with Ginga ($\Gamma\simeq 1.8$, Nandra \& Pounds
1994). No spectral flattening at high energies (above 10 keV), which
is usually found in Seyfert 1 nuclei and is considered to be due to
reflection from cold matter, was found (Iwasawa et al 1995; Smith \&
Done 1996). The profile of the broad iron K$\alpha $ emission peaks at
around 6.8 keV in the first ASCA spectrum taken in 1993 (Iwasawa et al
1996b). The iron line has a relatively large equivalent width, peaks
at an energy higher than 6.4 keV and the lack of a high energy hump
suggest X-ray reflection occurring from highly ionized disk rather
than a cold disk.

\section{Observation and data reduction}


\begin{table*}
\begin{center}
\caption{
  X-ray observations of IRAS18325--5926. The observation date given in
  Epoch column is the starting date of respective observation.
  Duration of each observation is given in unit of hour. The exposure
  time of each observation is useful time left after data selection.
}
\begin{tabular}{lcccc}
Epoch & Duration & Satellite & Exposure & F(2-10keV)\\
 & hr & & ks & $10^{-11}$\ergpspsqcm \\[5pt]
1989 May 17 & 11 & Ginga & 8.7 & 2.6 \\
1993 Sep 11 & 28 & ASCA & 36 & 1.3 \\   
1997 Mar 27 & 139 & ASCA & 243 & 1.9 \\
1997 Dec & 61 & RXTE & 131 & 2.4 \\
1998 Feb & 67 & RXTE & 134 & 1.8 \\
2000 Mar 31 & 75 & BeppoSAX & 115 & 2.0 \\
2001 Mar 05 & 33 & XMM & 120 & 1.2 \\
\end{tabular}
\end{center}
\end{table*}

The X-ray data used for the analysis in this paper were obtained from
Ginga, ASCA, RXTE, and BeppoSAX. The log of these observations
together with a most recent XMM-Newton observation is summarised in
Table 1. With the galaxy redshift $z=0.0198$ and \H0 = 70 \kmpspMpc,
flux of $1\times 10^{11}$\ergpspsqcm\ corresponds to luminosity of
$\approx 0.9\times 10^{43}$\ergps. Results from the Ginga and ASCA
observations have been published previously (Awaki et al 1991, Iwasawa
et al 1995, 1996, 1998; Smith \& Done 1996). The Ginga LAC spectrum is
the one analysed by Iwasawa et al (1995). Details of the data
reduction for the longer ASCA observation in 1997 can be found in
Iwasawa et al (1998). Results and details of the XMM-Newton
observation will be published elsewhere (Iwasawa et al 2003 in prep.).

The ASCA data obtained from the shorter observation in 1993 (reported
in Iwasawa et al 1996) have been reduced using the latest calibration.
In this ASCA observation, due to the pointed position of the
telescope, the source photons are spread over the four CCD chips in the SIS
detector, and non-negligible fraction of the total photons were lost
in the inter-chip gaps, which requires correction. Also the useful
exposure time of the SIS is shorter than that of the GIS.  To avoid
unnecessary complications, we only use the GIS data from this
observation. As a result of the updated calibration, the
averaged 2--10 keV flux is larger than that previously reported.

The duration of the observations ranges between 11 hr and 139 hr. The
spectral resolution in FWHM of the spectrometers at the Fe K band is
approximately 1 keV for the Ginga LAC and RXTE PCA, $\sim 500$ eV for
the ASCA GIS and BeppoSAX MECS, $\sim 130$ eV for the ASCA SIS
(for the 1993 observation; $\sim 250$ eV for the 1997 observation due
to detector degradation) and XMM-Newton pn camera. The mean 2--10
keV flux for each observation is given in Table 1. The lowest mean
flux was recorded during the latest XMM-Newton observation in 2001,
although the highest flux during the observation was twice as high as
the mean value.

Data reduction of two previously unpublished RXTE and one BeppoSAX
observations are described below. 

\subsection{RXTE PCA data}

The two RXTE observations were carried out on 1997 Dec 25--27, and
1998 Feb 21--24. The energy spectra from the two RXTE observations
were reduced as follows.  The PCA spectra were extracted only from the
top Xenon layer using the {\sc ftools 5.2} software.  Data from all
PCUs are combined, after weighting according to exposure to improve
signal-to-noise.

Good time intervals were selected to ensure stable pointing, and
exclude periods of South Atlantic Anomaly (SAA) passage. Elevation is
restricted to be more than $10^\circ$ above the Earth's limb.  We also
filter out electron contamination events. The net exposure
times for the two PCA spectra are 131 ks for the 1997 December
observation and 134 ks for the 1998 February observation.

We generate background data using {\sc pcabackest v3.0} in order  to
estimate the internal background caused by interactions between the
radiation/particles and the detector/spacecraft at the time of
observation.  This is done by matching the conditions of observations
with those in various model files.  The model files used are the
latest
\footnote[2]{{http://lheawww.gsfc.nasa.gov/users/craigm/pca-bkg/bkg-users.html}}~`CM'
background models (a refinement of the L7-240 models) which are
intended for application to faint sources with count
rate less than 100 cts/sec. The PCA response matrix for the {\it
RXTE} data set was created using {\sc pcarsp~v8.0}.  Background models
and response matrices are representative of the most up-to-date PCA
calibrations.

Despite the improved background estimate, there are still features
above 16 keV where the signal to noise ratio is low. The data analyzed
below are limited to the energy range of 3--16 keV.

\subsection{BeppoSAX data}

IRAS 18325--5926 was observed with BeppoSAX during the period between
2000 March 31 and 2000 April 3. Spectral data are available from three
detectors, LECS, MECS and PDS, which are sensitive and calibrated for
a scientific analysis to X-rays in the energy ranges of 0.1--4 keV,
2--10 keV and 14--200 keV, respectively.  The event files for the MECS
and LECS data were provided by the SAX Data Center (SDC). Events from
the two MECS detectors (MECS2 and MECS3) have been merged and the MECS
spectrum was extracted from the merged event file. The PDS spectrum
which has been reduced and corrected for background by SDC is used for
our analysis. The count rates in the respective detectors for net
exposure times are 0.087 \cps\ for 39 ks (LECS), 0.23 \cps\ for 115 ks
(MECS) and 0.13 \cps\ for 59 ks (PDS). The X-ray flux quoted is
obtained from the MECS. The relative normalization ratio of the LECS
to the MECS was found to be $\simeq 0.7$ from fitting, while that of
the PDS is fixed at 0.86 in the spectral analysis.

\section{Results}

\subsection{Light curves from RXTE and BeppoSAX}

Light curves in the 2--10 keV band from the RXTE and BeppoSAX
observations are shown in Fig. 1. The RXTE light curves are obtained
from the five PCU while the BeppoSAX light curve is from the MECS.
Although there is a hint of X-ray flux peaking every 40--50 ks
intervals during the RXTE 1998 observation, no significant periodic
signals are found in these light curves, contrary to the nine cycles
of 16-hr quasi-periodic modulations found in the five-day long ASCA
observation in 1997 (Iwasawa et al 1998).

Count rate ratios between three energy bands (2--4 keV, 4--7 keV, and
7--12 keV) have been examined, using the RXTE data. There are
occasional deviations from the mean value but no correlated ratio
changes with total flux were found. As a whole, the data are
consistent with no variations in X-ray colour.


\begin{figure}
\centerline{\includegraphics[width=0.3\textwidth,angle=270,
    keepaspectratio='true']{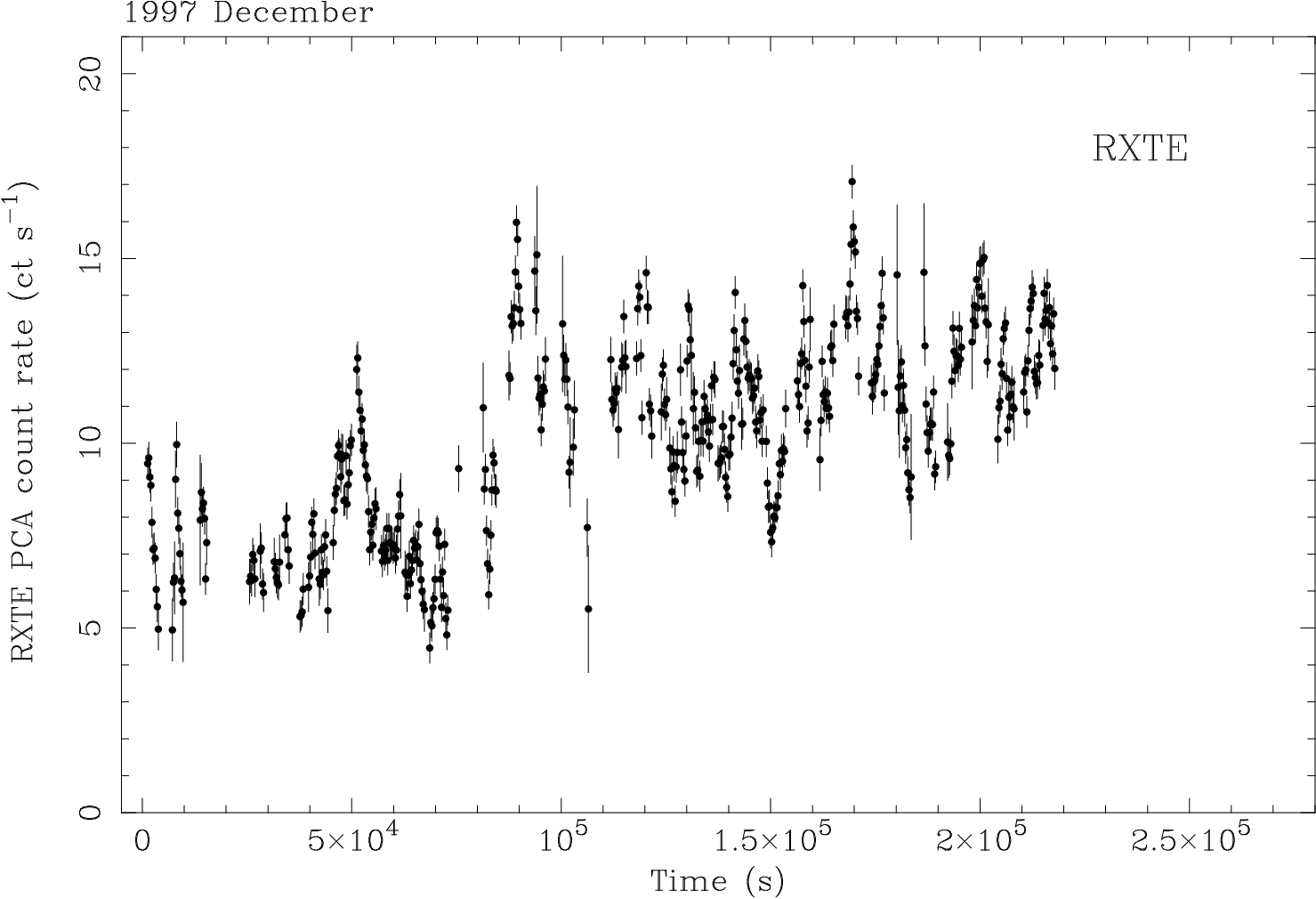}} 
\centerline{\includegraphics[width=0.3\textwidth,angle=270,
    keepaspectratio='true']{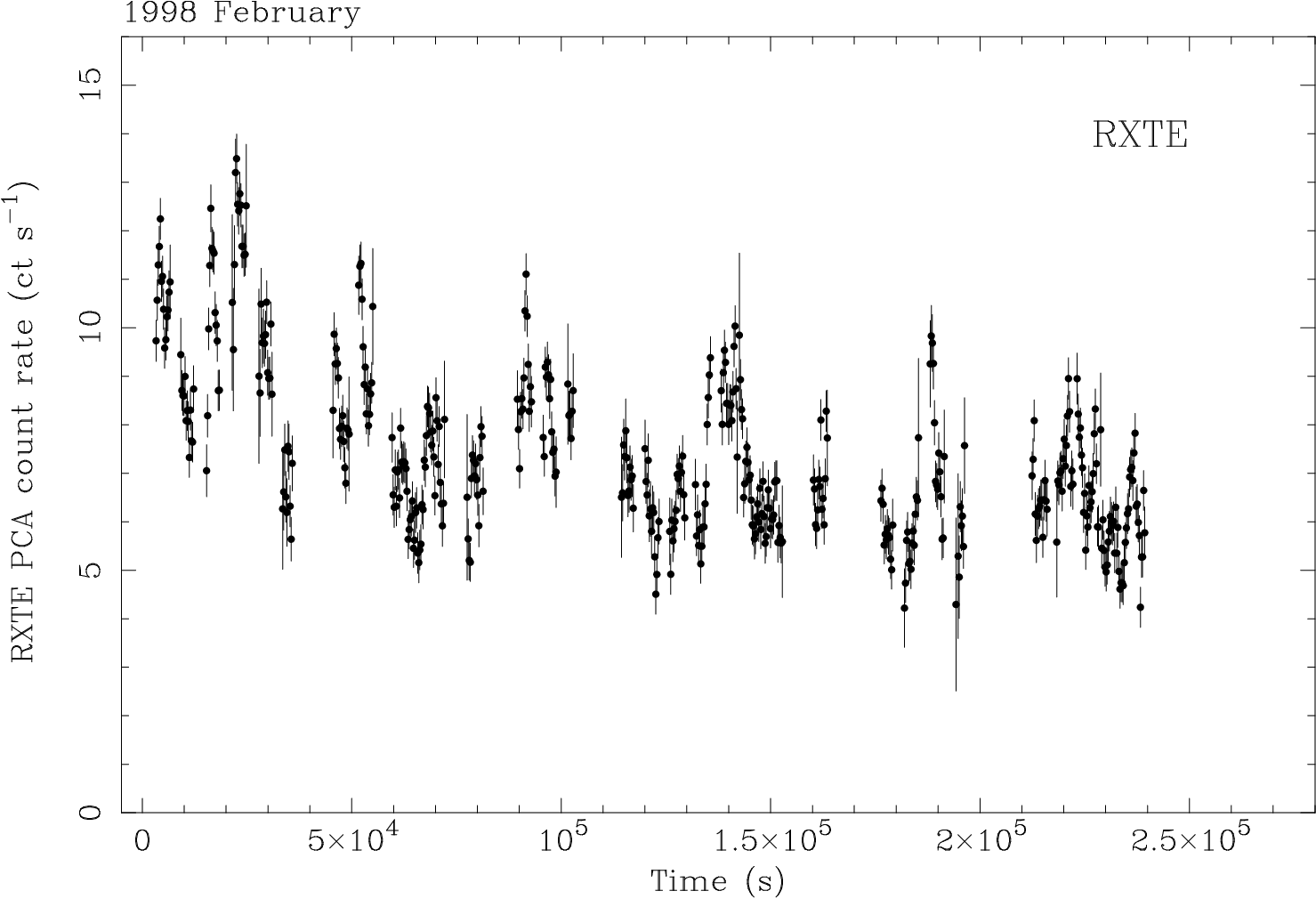}} 
\centerline{\includegraphics[width=0.3\textwidth,angle=270,
    keepaspectratio='true']{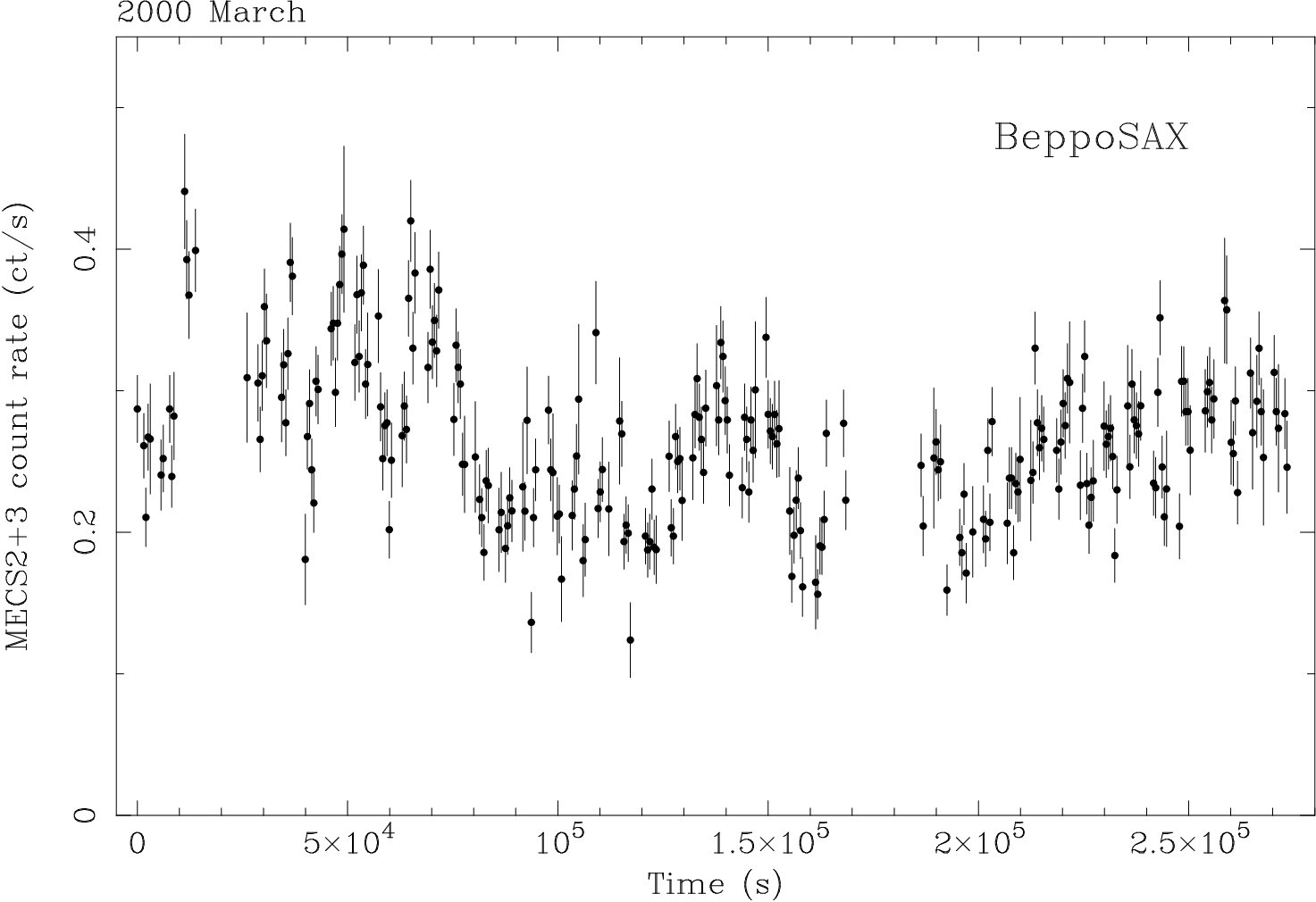}} 
\caption{
  The 2--10 keV X-ray light curves of IRAS 18325--5926 during the two
  RXTE observations in 1997 and 1998 and the BeppoSAX observation in
  1999.}
\end{figure}

\subsection{Steep spectral slope}


\begin{figure}
\centerline{\includegraphics[width=0.36\textwidth,angle=270,
    keepaspectratio='true']{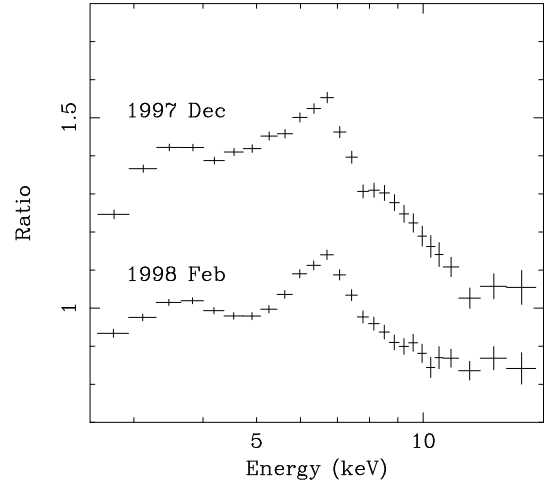}} 
\caption{
  Plot of the ratio of the RXTE spectra taken in 1997 Dec and 1998 Feb
  to the absorbed power-law model with $\Gamma = 2$ and \nH $=1\times
  10^{22}$\psqcm. The normalization of the power-law is adjusted to
  match the 1998 Feb spectrum at around 3--5 keV. Spectral steepening and
  a significant edge-like feature above 10 keV are recognised in the
  1997 Dec observation.}
\end{figure}

An earlier measurement of the spectral slope for the 2--18 keV
spectrum from the Ginga LAC indicated a relatively steep power-law
slope ($\Gamma\sim 2.2$). No significant evidence for
spectral hardening above 10 keV (Iwasawa et al 1995; Smith \& Done
1996), which are typically found in normal Seyfert 1 galaxies (Nandra
\& Pounds 1994), was found. The spectra obtained from the RXTE PCA and the
BeppoSAX MECS/PDS confirmed these results and revealed more spectral
complexity. Here we discuss the RXTE and BeppoSAX spectra.

Fig. 2 shows plots of the ratios of the RXTE PCA spectra against an
absorbed power-law with photon index of $\Gamma = 2.0$ and an excess
absorption column density above the Galactic value (\nH $= 7.4\times
10^{20}$\psqcm, Dickey \& Lockman 1990) \nH $ = 1\times
10^{22}$\psqcm. The normalization of the power-law is adjusted so
that the 3--5 keV range matches the 1998 February data. This absorbed
power-law was chosen only for the purpose of displaying spectral features. 

It is clear from Fig. 2 that the continuum slopes of both spectra are
steeper than $\Gamma = 2.0$. A naive estimate of a power-law slope for
the continuum in the two respective spectra are $\Gamma\sim 2.2$,
although a simple power-law fit even with a gaussian for the Fe
K$\alpha $ does not provide a good agreement with the data due to more
complex spectral features, as described in the following section. Fig.
2 also shows that the spectrum during the 1997 Dec observation is
steeper than that during the 1998 February observation: the observed
flux in the 1997 Dec data is brighter by $\sim 40$ per cent at 3 keV
than the 1998 Feb data but only by $\sim 20$ per cent at 15 keV.

\subsection{Possible high energy roll-over}


\begin{figure}
\centerline{\includegraphics[width=0.35\textwidth,angle=270,
    keepaspectratio='true']{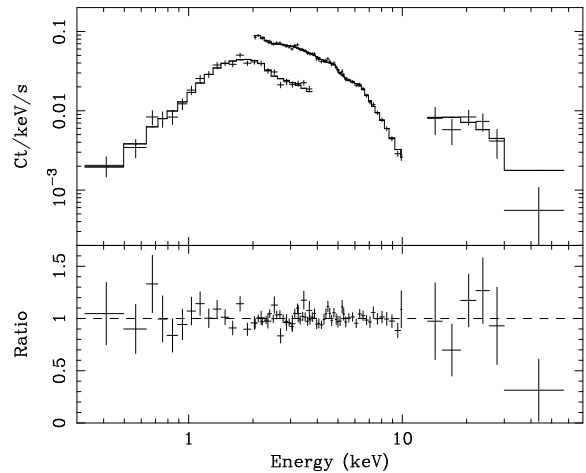}} 
\caption{
  The BeppoSAX spectrum of IRAS 18325--5926. The LECS (0.3--4 keV),
  MECS (2--10 keV) and PDS (14--60 keV) data are plotted with the
  model (solid-line histogram) consisting of a power-law and a
  gaussian iron K$\alpha $ line modified by a partially covering cold
  absorber at the galaxy's redshift and Galactic absorption, which
  best-fits the LECS and MECS data. Extrapolating the model to
  higher energies matches the PDS data up to 30 keV but overestimates
  significantly above the energy, as shown in the residual in the form
  of ratio of the data and the model. }
\end{figure}

The broad-band X-ray spectrum of IRAS 18325--5926 is investigated
using data from the three detectors on-board BeppoSAX (Fig. 3). As
noted in the previous ASCA and ROSAT PSPC observations, there is an
excess component below 1 keV which appears to be less variable than
the higher energy emission (Iwasawa et al 1996). This has been
confirmed with the longer ASCA observation.  Although the origin of
this soft excess component is unclear, we tentatively model this
component using a partial covering absorber to a power-law for
describing the soft X-ray spectrum obtained from the BeppoSAX LECS.

This partially absorbed power-law plus a gaussian for the Fe K$\alpha
$ emission provides a good fit to the MECS and LECS data in the
0.3--10 keV band with $\chi^2 = 110.5$ for 89 degrees of freedom. The
Galactic absorption \nH $= 7.4\times 10^{20}$\psqcm\ is included in the
fit.  Best-fitting spectral parameters for the continuum are $\Gamma =
2.13^{+0.06}_{-0.07}$, \nH $= 1.3^{+0.1}_{-0.2}\times 10^{22}$\psqcm,
and the covering fraction of the absorber, $f_{\rm C} =
0.95^{+0.01}_{-0.01}$; hereafter errors quoted to spectral parameters
are of the 90 per cent confidence limit for one parameter of interest
unless stated otherwise.

Extrapolating this continuum model agrees with the PDS data up to 30
keV, but overestimates the data at energies above 30 keV. The PDS data
above 50 keV is essentially at the noise level. The detected count rate in
the 30--50 keV band are factor of 4 ($\geq 1.8$) below the
extrapolation of the power-law continuum best-fitting the MECS data,
and its deviation is $2\sigma$. 

Apparently the high energy roll-over is too steep to be an exponential
cut-off ($\exp(-E/E_{\rm c}$) where $E_{\rm c}$ is a cut-off energy in
keV) to the power-law. It is better described by a sharp cut-off at
$29^{+16}_{-15}$ keV with an e-folding energy of 6 keV ($\exp[(E_{\rm
  c}-E)/E_{\rm f}]$ for $E>E_{\rm c}$), which is however, not
constrained very well ($E_{\rm f}\leq 100$ keV). Background
subtraction at this high-energy-end introduces an additional
uncertainty, especially given the low count rate: a subtle
oversubtraction of the intrumental line at around 50 keV in the
background could cause a sharp decline in the source spectrum, such as
that seen in our data. The HEXTE data from the RXTE observations do not
have sufficient quality to provide any useful constraints on the
spectral shape in this energy band. We therefore regard this high
energy deficit as a tentative detection of a spectral roll-over at 30
keV. This roll-over, however, would have little effect on the
measurements of the strength of reflection using the data up to $\sim
20$ keV such as Ginga LAC and RXTE PCA data.

As discussed in the following section, the X-ray spectrum can be
explained by a model with reflection from an ionized disk. The
reflection spectrum produces a spectral bump around 30 keV due to
Compton down-scattering, which could partly explain the observed drop
of X-ray flux above 30 keV (see e.g., Barrio, Done \& Nayakshin 2003).
As noted by e.g., Stern et al (1995) and Petrucci et al (2001), the
temperature of the Comptonizing corona does not necessarily inferred
from an energy of the power-law cutoff. We then fit the BeppoSAX data
with the Comptonization model {\tt compps} by Poutanen \& Svennson
(1996), combined with the ionized reflection model {\tt xion} by S.
Nayakshin (the details of which will be described in Section 3.7).
Note that the {\tt xion} code assumes a power-law with a high energy
cut-off as an illuminating source rather than the Comptonized
continuum, which should, however, little effect on the fit.  The
fittings were performed assuming either spherical or cylinder
geometries, which give similar quality of fit with a coronal
temperature of $kT_{\rm e}$ = 35--40 keV and optical depth of
$\tau\simeq 2$. The statistical uncertainty of the drived temperature
is the order of $\pm 5$ keV, but the systematic error mentioned above
should dominate the uncertainty.

\subsection{Emission lines and absorption edge}

Here, we present a spectral analysis of the data from Ginga, RXTE,
BeppoSAX, ASCA and XMM-Newton, using a simple phenomenological model.
The fitted model consists of an absorbed power-law and a gaussian for
the broad Fe K$\alpha $ line. Results of spectral fits are shown in
Table 2. Variability and correlations of some key spectral parameters
will be discussed in the next subsection.

The 1997 Dec spectrum shows a significant deficit against a power-law
between 10 keV and 15 keV, of which shape can be approximated by an
absorption edge at $10.7^{+0.4}_{-0.4}$ keV with optical depth $\tau =
0.18^{+0.05}_{-0.06}$ (Fig. 2). No strong absorption edge is expected
at this energy. An immediate interpretation of this feature is a
blueshifted Fe K absorption edge, but an alternative origin of this feature is
discussed in Section 3.7. Such an edge-like feature is not
significantly detected in any other spectra (e.g., the 90 per cent
upper limit on the optical depth of an absorption edge at the same
energy in the 1998 Feb RXTE spectrum is $\tau\leq 0.12$).

In addition to the prominent Fe K$\alpha $ emission, there is a weaker
excess at $\sim 3.4$ keV in the RXTE spectra (Fig. 2). This features
is also present in the Ginga LAC spectrum with lower signal-to-noise
ratio and the good quality XMM-Newton spectrum. It can be identified
with Radiative Recombination Continuum (RRC) of SXVI in the context of
ionized reflection model discussed below. Although uncertainties in
calibration of the detector response at lowest energies are
suspected for the RXTE PCA, since the excess is at $3\sigma $ level in
the two RXTE spectra and the XMM-Newton spectrum has detected the same
feature clearly (Iwasawa et al 2003 in prep.), the presence of the
feature is highly probable.


\begin{table*}
\begin{center}
\caption{
  Spectral fits to the Ginga, RXTE, ASCA, BeppoSAX, and XMM-Newton pn
  spectra. The fitted model consists of an absorbed power-law with a
  gaussian for Fe K$\alpha$. The line centroid energy has been
  corrected for the galaxy redshift ($z=0.0198$). It should be noted
  that the line centroid does not necessarily coincide with the peak of the
  emission line profile when a skewed profile is fitted, particularly
  in a low resolution spectrum, by a gaussian which has as symmetric
  shape. $^{\star}$The best-fit spectral parameters given for the 1997
  Dec RXTE spectrum are those obtained when an absorption edge is also
  included to describe the 10--15 keV deficit with $\chi^2 = 44.64$
  for 28 degrees of freedom (see text for details), but the $\chi^2$
  value given in this table is for the fit without the edge model for a
  comparison with the other fits. }
\begin{tabular}{lcccccccc}
Data & Band & $\Gamma $ & \nH & $E_{\rm Fe}$ & $\sigma_{\rm Fe}$ & $I_{\rm Fe}$
& $EW_{\rm Fe}$ & $\chi^2$/dof \\
& keV & & $10^{22}$\psqcm & keV & keV & $10^{-5}$\phpspsqcm & eV & \\[5pt]
Ginga & 2-18 & $2.26^{+0.05}_{-0.06}$ & $1.42^{+0.10}_{-0.21}$ &
$6.38^{+0.22}_{-0.12}$ & $0.64^{+0.17}_{-0.23}$ & $11.1^{+4.3}_{-3.1}$ &
411 & 14.16/21 \\
ASCA93 & 2-10 & $2.26^{+0.20}_{-0.15}$ & $1.47^{+0.44}_{-0.38}$ &
$6.54^{+0.17}_{-0.20}$ & $0.59^{+0.20}_{-0.19}$ & $7.64^{+6.06}_{-2.38}$ &
626 & 422.1/418 \\
ASCA97 & 2-10 & $2.00^{+0.03}_{-0.05}$ & $1.04^{+0.12}_{-0.12}$ &
$6.59^{+0.11}_{-0.11}$ & $0.58^{+0.16}_{-0.17}$ & $5.46^{+1.62}_{-1.53}$ &
268 & 1856.4/1802 \\
RXTE97 & 3-16 & $2.24^{+0.02}_{-0.03}$ & $2.37^{+0.29}_{-0.29}$ &
$6.53^{+0.08}_{-0.09}$ & $0.46^{+0.16}_{-0.10}$ & $6.52^{+1.29}_{-1.07}$ &
244 & 74.00/30$^{\star}$ \\
RXTE98 & 3-16 & $2.17^{+0.03}_{-0.02}$ & $1.62^{+0.29}_{-0.12}$ &
$6.65^{+0.07}_{-0.06}$ & $0.41^{+0.16}_{-0.04}$ & $5.58^{+1.40}_{-1.27}$ &
312 & 42.82/30 \\
BeppoSAX & 2-10 & $2.19^{+0.11}_{-0.08}$ & $1.52^{+0.22}_{-0.23}$ &
$6.50^{+0.24}_{-0.34}$ & $0.42^{+0.88}_{-0.42}$ & $4.47^{+8.13}_{-2.60}$ &
201 & 77.53/64 \\
XMM & 2-11 & $2.06^{+0.03}_{-0.03}$ & $1.24^{+0.08}_{-0.11}$ &
$6.61^{+0.07}_{-0.06}$ & $0.43^{+0.04}_{-0.11}$ & $2.99^{+0.43}_{-0.57}$ &
242 & 1119.0/1125 \\
\end{tabular}
\end{center}
\end{table*}

\subsection{Variable iron K line emission}



\begin{figure*}
\centerline{\includegraphics[width=0.37\textwidth,angle=270,
    keepaspectratio='true']{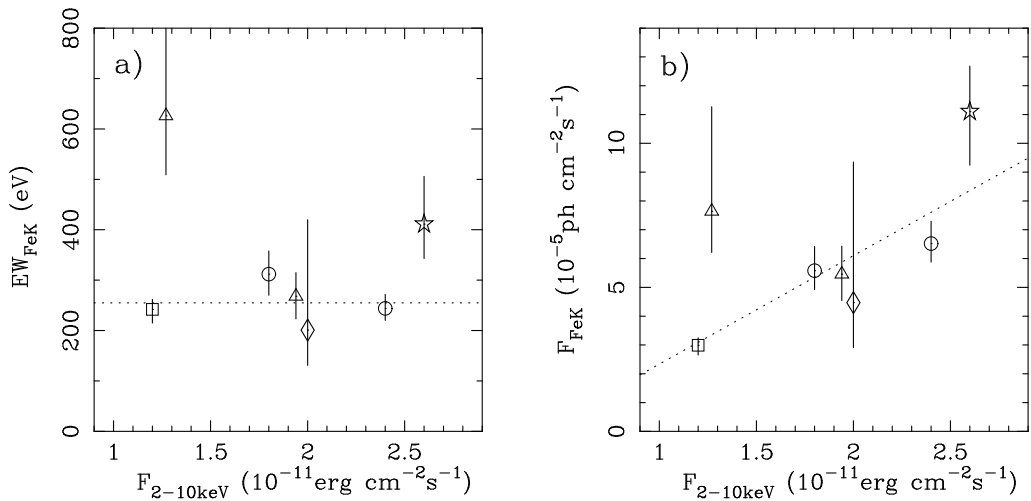}} 
\caption{
  a) Plot of equivalent width (EW) of the Fe K$\alpha $ line, derived
  by fitting a gaussian, against the 2--10 keV flux for the
  observations with Ginga (star), ASCA (triangles), RXTE (circles),
  BeppoSAX (diamond) and XMM-Newton (square). The dotted line shows a
  fit with a constant value ($EW_{\rm FeK} = 255^{+24}_{-22}$ eV) for
  the data except the ASCA93 observation with $\chi^2 = 4.9$ for 5
  degrees of freedom. b) Plot of Fe K line flux against the 2--10 keV
  flux (symbols represents the same observations as Fig. 5a).  Error
  bars represents 1 $\sigma$ errors. The dotted line shows a linear
  correlation of $F_{\rm FeK} = -1.42^{+1.32}_{-1.32} +
  3.67^{+0.89}_{-0.89}F_{\rm 2-10keV}$ where $F_{\rm FeK}$ and $F_{\rm
    2-10keV}$ are Fe K$\alpha $ line flux and the 2--10 keV flux in
  the unit used in the plot ($\chi^2 = 5.2$ for 4 degrees of freedom).
  The fit excludes the data point from the ASCA93 observation.}
\end{figure*}

The 2--10 keV flux averaged over each observation was observed to vary
by a factor of 3 between the seven observations analyzed above (Table
1). Any response of the iron K line to the continuum is of great
interest in the context of a reflection model for producing the iron
line emission, although, with only seven data points, it is premature
to discuss the reality of a correlation on statistical ground.

Apart from the first ASCA observation in 1993 when the line EW is
unusually large, a correlation between the line and continuum flux
appears to exist (Fig. 4). Since there might be uncertainties in
absolute flux calibration between the instruments, the EW of the lines,
which are independent of the cross-calibration errors, are plotted
against the measured 2--10 keV flux (see also Fig. 5) along with the
plot of the line fluxes. \footnote[3]{ASCA and BeppoSAX are in
  agreement within 3 per cent in absolute flux in a simultaneous
  observation of 3C273 (ASCA GOF Calibration Memo 06/07/00). With the
  latest calibration, the RXTE measurement should be in a reasonable
  agreement (e.g., $\sim 10$ per cent) with the above two, although
  fluctuation of the diffuse X-ray background (XRB) is a major
  uncertainty for a X-ray source with brightness of IRAS 18325--5926.
  The Ginga flux is less affected by the XRB problem, because a local
  background data was taken.} One could also claim that the line flux
is continuously decreasing regardless of the continuum luminosity
(Table 2).

The line shape remains roughly the same between the observations. The
means of the line centroid energy and line width are 6.59 keV and 0.47
keV (these values are weighted means). The iron line shape in the form
of ratio against the best-fitting power-law continuum for all spectra
are shown in Fig. 5.


\begin{figure*}
\hbox{\hspace{5mm}{\includegraphics[width=0.32\textwidth,angle=270,
    keepaspectratio='true']{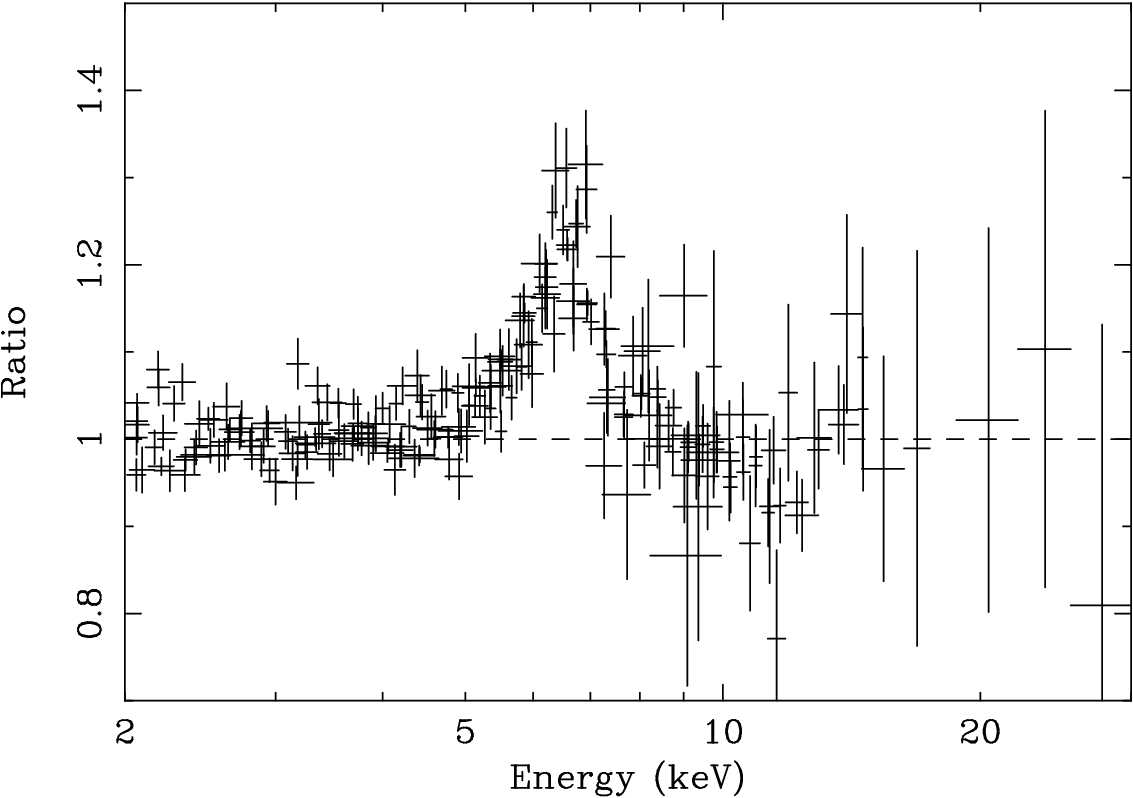}}\hspace{5mm}
{\includegraphics[width=0.32\textwidth,angle=270,keepaspectratio='true']{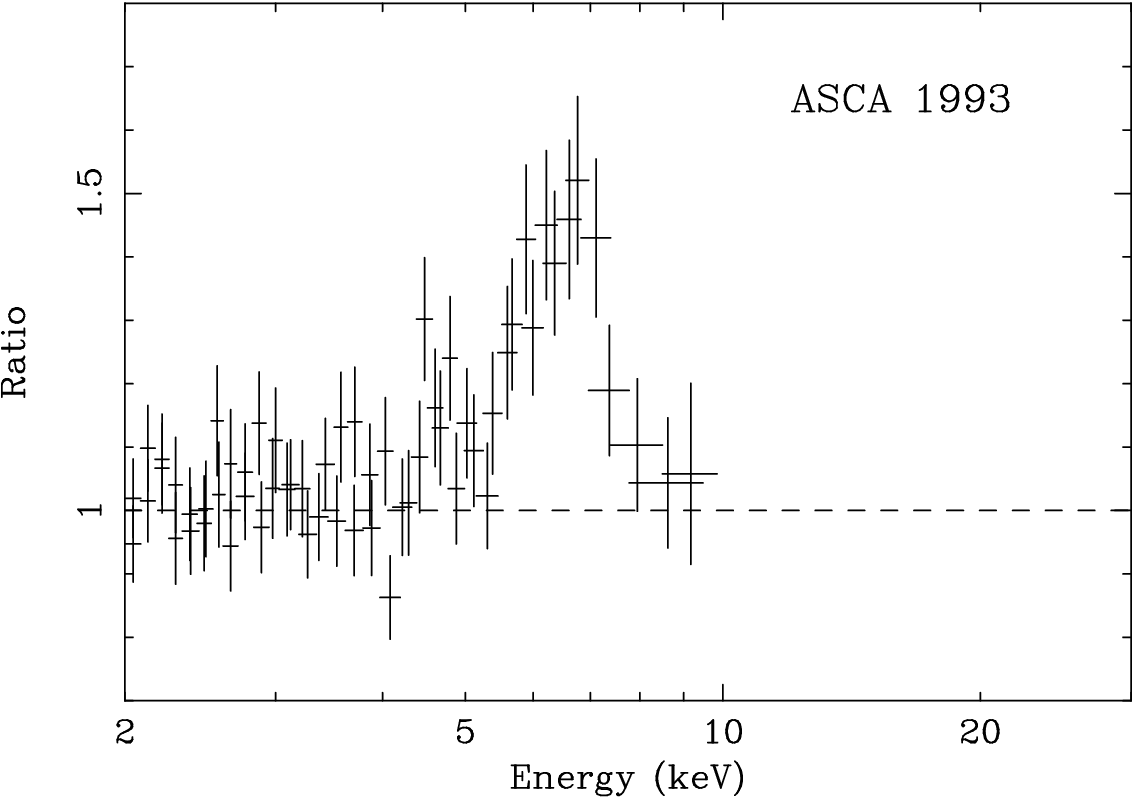}}}
\caption{
  Left: The Fe K$\alpha $ line profiles from all the observations
  except for the ASCA 1993 observation which is plotted separately in
  the right panel in ratio relative to respective best-fit power-law
  continua. The data plotted here were obtained from detectors with a
  broad range of spectral resolution (from a proportional counter to
  an X-ray CCD). However, the measured line widths (see Table 2) are
  comparable with the worst resolution of the proportional counters,
  which means that the overall shape of the lines can be compared
  regardless the resolving power of the detector. Right: The same plot
  obtained from the ASCA 1993 observation (Iwasawa et al 1996).  Note
  the difference in y-axis scale. The ASCA 1993 line is approximately
  twice as strong as the others.}
\end{figure*}

\subsection{Evidence for reflection from an ionized disk}

\begin{table*}
\begin{center}
\caption{
  Spectral fitting with reflection spectra from a photoionized disk
  in hydrostatic balance by S. Nayakshin. The ``magnetic flare''
  geometry is chosen. The illuminating power-law source is assumed to
  have a high energy cut-off at 60 keV and variable photon index
  $\Gamma$. The dimensionless accretion rate (relative to the
  Eddington limit) is set at 0.5. The inclination angle ($i$), inner
  radius (\Rin), outer radius (\Rout), and radial emissivity index
  ($\alpha $; emissivity $\propto r^{-\alpha }$) of the accretion disk
  are assumed to be cos $i = 0.9$, \Rin\ = 3\Rs, \Rout\ = 100\Rs, and
  $\alpha = 2$, respectively, for the all spectra apart from the two
  RXTE and XMM-Newton datasets for which \Rin = 3\Rs\ was found to be
  outside the 90 per cent confidence limit and allowed to vary. \Rs\ 
  is Schwarzschild radius ($= 2 r_{\rm g}$). The metallicity of the
  disk material is assumed to be solar value. Relativistic blurring
  for the Schwarzschild metric (Fabian et al 1989) is applied. The
  luminosity ratio of the X-ray illuminating source and thermal
  emission from the disk, $l_{\rm x}/l_{\rm d}$, is left as the only
  free parameter in the ionized disk model. A sum of the illuminating
  power-law and reflection, which is absorbed by a cold absorbing
  column \nH, is compared with the data. The energy band used for the
  spectral fittings in each dataset is as the same as that is given in
  Table 2.}
\begin{tabular}{lccccc}
Data & $\Gamma $ & \nH & $l_{\rm x}/l_{\rm d}$ & \Rin & $\chi^2$/dof \\
& & $10^{22}$\psqcm & & \Rs & \\[5pt]
Ginga & $2.20^{+0.08}_{-0.06}$ & $1.65^{+0.25}_{-0.21}$ & $\leq 0.64$ &
3 & 12.5/23 \\
ASCA93 & $2.17^{+0.09}_{-0.11}$ & $1.69^{+0.28}_{-0.35}$ & 
$0.08^{+0.13}_{-0.05}$ & 3 & 423.8/420 \\
ASCA97 & $1.96^{+0.05}_{-0.04}$ & $1.20^{+0.05}_{-0.05}$ & 
$1.0^{+0.4}_{-0.4}$ & 3 & 1867.6/1804 \\
RXTE97 & $2.28^{+0.03}_{-0.04}$ & $3.00^{+0.22}_{-0.30}$ & 
$0.35^{+0.21}_{-0.14}$ & $18^{+5}_{-8}$ & 48.6/31 \\
RXTE98 & $2.21^{+0.04}_{-0.05}$ & $2.36^{+0.22}_{-0.34}$ & 
$\leq 0.2$ & $34^{+21}_{-9}$ & 39.9/31 \\
BeppoSAX & $2.16^{+0.08}_{-0.06}$ & $1.68^{+0.24}_{-0.22}$ & 
$1.0^{+1.0}_{-0.6}$ & 3 & 76.8/66 \\
XMM & $2.03^{+0.03}_{-0.01}$ & $1.36^{+0.08}_{-0.05}$ & 
$0.28^{+0.19}_{-0.10}$ & $19^{+26}_{-5}$ & 1130.4/1126 \\
\end{tabular}
\end{center}
\end{table*}

One notable feature in the X-ray spectrum of IRAS 18325--5926 is the
broad, Fe K$\alpha $ emission peaking at energies significantly higher
than 6.4 keV, at which most Fe K$\alpha $ lines of Seyfert 1 galaxies
are found (e.g., Nandra et al 1997b). To interpret the iron line
feature, reflection from a highly ionized disk is favoured for the
following reasons. 

The shape of the iron $K\alpha $ line can be explained by broadened
FeXXV at 6.7 keV. A very high EW ($\sim 600$ eV, see Table 2 and
Iwasawa et al 1996) was observed in the spectrum of the ASCA 1993
observation. The high fluorescence yield of FeXXV (Matt et al 1993)
can explain such a large EW. A bump at around 3.4 keV can be
identified with the RRC of SXVI (Section 3.4), which is expected in a
reflection spectrum with strong FeXXV.

No significant high energy hump above 10 keV is observed (Iwasawa et
al 1995; Smith \& Done 1996). Compared to the case of cold reflection
in which a high energy hump due to Compton down-scattering stands out
when the incident power-law and reflection spectrum are added (e.g.,
George \& Fabian 1991), in ionized reflection, strong reflection takes
place also at lower energies where photons that entered and are
reflected back in the ionized matter are no longer subject of
photoelectric absorption, resulting in the Compton down-scattered hump
being less pronounced because of a smaller contrast between low and
high energy ranges (e.g., Ross \& Fabian 1993).

A recent study of a photoionized disk in hydrostatic equilibrium by
Nayakshin and collaborators (Nayakshin et al 2000) predicts that AGN
with a steep continuum slope ($\Gamma>2$) tend to show distinctive
``ionized disk'' signatures, e.g., strong FeXXV+FeXXVI $K\alpha$
emission, as predicted in the constant density models (e.g., Ross \&
Fabian 1993; Ross, Fabian \& Young 1999; Ballantyne, Ross \& Fabian
2001). This is because the softer ionizing radiation causes the
Compton temperature of the ionized surface of the disk to remain cool
so that the region is not completely ionized. A low cut-off energy of
the incident power-law also helps for the same reason. These
conditions fit the photon index ($\Gamma\simeq 2.2$, Section 3.2) and
the possible spectral roll-over at 30 keV (Section 3.3) observed in
IRAS 18325--5926.

\subsection{Fitting the ionized reflection model}


\begin{figure}
\centerline{\includegraphics[width=0.3\textwidth,angle=270,
    keepaspectratio='true']{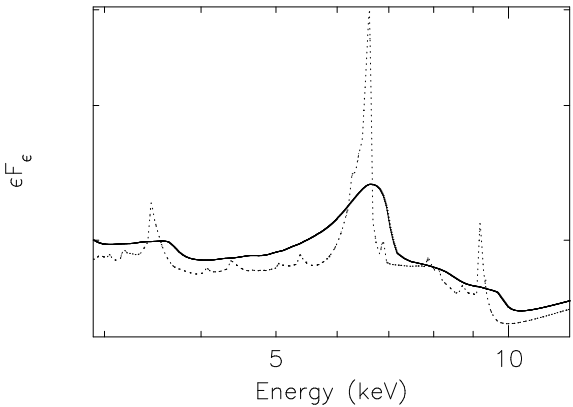}} 
\caption{
  An example of the ionized reflection spectra with (solid line) and
  without (dotted line) relativistic blurring. This particular example
  is for the best-fit model for the 1997 RXTE data.}
\end{figure}


\begin{figure}
\centerline{\includegraphics[width=0.6\textwidth,angle=270,
    keepaspectratio='true']{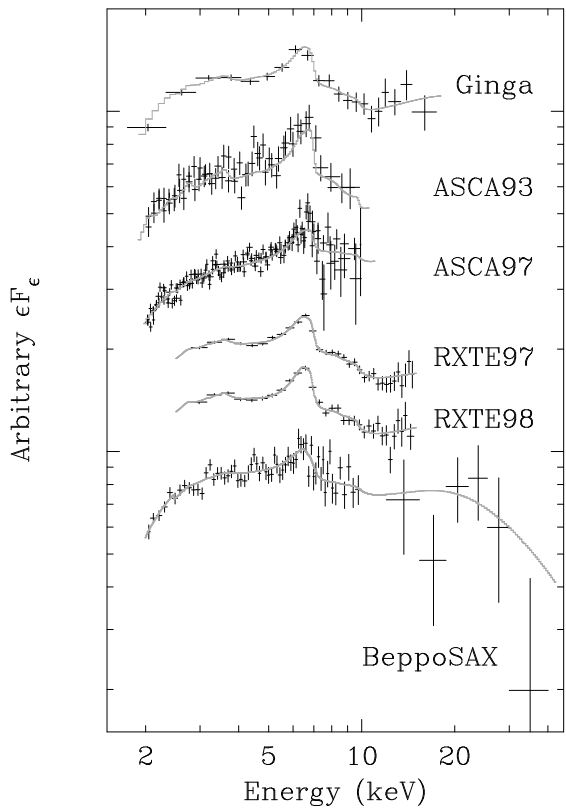}} 
\caption{
  Spectra fitted with ionized reflection model (as shown in grey
  solid-lines). The data are shifted along y-axis for clarity and
  plotted in the chronological order (top to bottom for old to new).}
\end{figure}

We have compared spectra expected from the ionized disk models with
the observed X-ray spectra of IRAS 18325--5926. The presence of the SXVI
RRC feature is in favour of using the reflection model from an ionized
disk computed by {\tt xion}, over the models by Ross \&
Fabian (1993) in which atomic features of sulphur have not been
included.  

The ionized reflection model gives good fits to all the seven
spectra. The iron line feature at 6--7 keV comprises multiple lines
but is dominated by FeXXV K$\alpha$ at 6.7 keV in these models. The
quality of the fits are comparable with or sometimes better than the
phenomenological model with an absorbed power-law and a gaussian given
in Table 2. The improvement in the ionized reflection fits comes from
explaining the SXVI RRC feature around 3.4 keV in the high
signal-to-noise ratio data from Ginga and RXTE. 

Sharp spectral features like Fe K$\alpha$ line are broadened to
some degree via Compton scatterings within the reflection layer when
the disk is highly ionized. However, the broadening by Compton
scattering is insufficient to explain the observed iron line profiles
and the data require significant relativistic broadening, expected
for the region close to a central black hole.

This relativistically blurred ionized reflection can partly explain
the puzzling edge-like feature at 10.6 keV in the 1997 Dec RXTE
spectrum: broadened FeXXV RRC, which would peak at 9.2 keV in the
absence of relativistic broadening, makes a shoulder dropping at
around 10 keV, mimicking an absorption edge at 10.5 keV (Fig. 6).
However, a shallow deficit still remains. Introducing an extra
absorption edge at $E_{\rm th}=11.4^{+0.7}_{-0.7}$ keV with optical
depth $\tau = 0.11$ improves the fit by $\Delta\chi^2 = 9.4$ from the
fitting with the reflection model alone given in Table 3.

In order to make a comparison between the datasets easier, we restrict
the number of free parameters to a minimum in the spectral fits. Apart
from the photon index ($\Gamma $) of the illuminating power-law, its
normalization and cold absorption column density (\nH), only the
luminosity ratio of the illuminating X-ray source and thermal emission
from the disk, $l_{\rm x}/l_{\rm d}$, is allowed to vary where
possible. This parameter basically defines the Compton temperature of
the ionized surface layer of the disk (Nayakshin \& Kallman 2001),
which controls the strength of the spectral features, e.g., iron
K$\alpha $ line emission. The ``magnetic flare'' geometry, in which
the disk is illuminated locally by individual flares (e.g., Haardt,
Maraschi \& Ghisellini 1997), is used. The power-law source is assumed
to have a cut-off energy at 60 keV. The dimensionless accretion rate
relative to the Eddington limit is set to 0.5. Fitting the individual
spectra yield values ranging in 0.3--0.7 for this parameter, but they
do not differ significantly each other within errors. Hence it is
fixed at 0.5 in the following fittings. This parameter defines the
ionization parameter of the disk, the above value means that the disk
is highly ionized to emit FeXXV. The inclination ($i$), inner radius,
outer radius and radial emissivity index ($\alpha $) of the disk are
assumed to be cos $i=0.9$, \Rin $=3$\Rs, \Rout $=100$\Rs, and $\alpha
= 2$, respectively, where \Rs\ is Schwarzschild radius (=2\rg).
However, \Rin\ is allowed to vary in some cases when 3 \Rs\ is found
out to be outside the 90 per cent confidence limit (for RXTE97, RXTE98
and XMM-Newton). This parameter is chosen to control the strength of
relativistic broadening, but it should be noted that the value
obtained from the fit does not necessarily mean the true inner
boundary of the disk, as it is coupled with the other parameters such
as the emissivity index, which is not known. The disk gas is assumed
to have Solar metallicity. The results of spectral fitting are shown
in Table 3 and Fig.  7. The ionized reflection models reproduce the
complex spectral features seen in the observed data well, including
the strengths of Fe K$\alpha $ and SXVI RRC.

As we found with the gaussian fits (Table 2 and Fig. 4b), the iron
line flux appears to correlate with the continuum flux, but the ASCA93
data shows a strong line despite the low continuum flux level, making
it an outlier of the correlation. The large EW of the iron line in the
ASCA 93 spectrum is explained by the reflection model with a small
value of $l_{\rm x}/l_{\rm d}$, because a cool (temperature of a few
hundreds eV) layer producing FeXXV becomes larger in this condition.
The plot of $l_{\rm x}/l_{\rm d}$ against iron K$\alpha $ line EW
(Fig. 8) shows this trend, that is, in general, small EW is associated
with large $l_{\rm x}/l_{\rm d}$, and vice versa.

\begin{figure}
\centerline{\includegraphics[width=0.35\textwidth,angle=270,
    keepaspectratio='true']{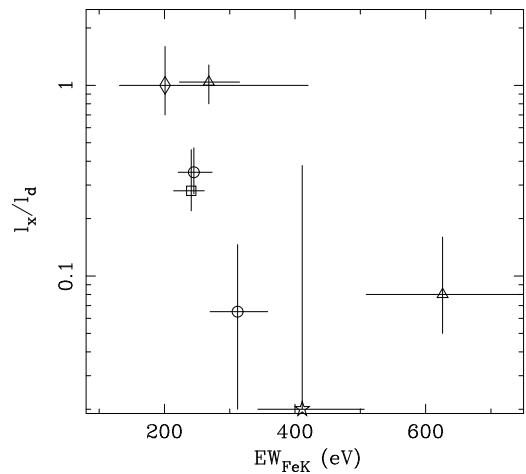}} 
\caption{
  The $l_{\rm x}/l_{\rm d}$ parameter in the ionized reflection model
  fits against Fe K$\alpha $ line equivalent width obtained from
  gaussian fits. Symbols are as in Fig. 4. Error bars represent
  $1\sigma$ errors.}
\end{figure}

\subsection{Iron line variability in the long ASCA observation}


\begin{figure}
\centerline{\includegraphics[width=0.32\textwidth,angle=270,
    keepaspectratio='true']{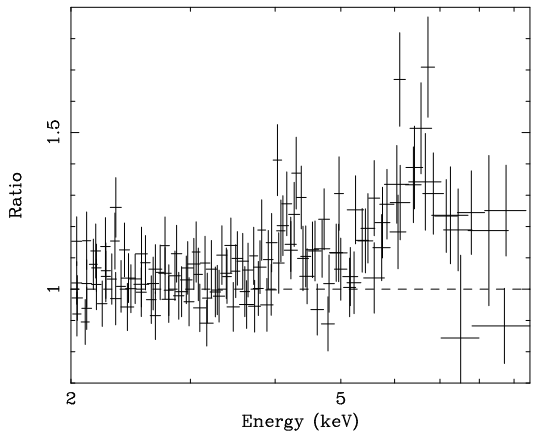}} 
\caption{
  The Fe K$\alpha $ line profile obtained from the last 20 hr of the
  1997 ASCA observation (the exposure time is $\sim 28$ ks). The data
  are from the four detectors and plotted in the form of ratio to the
  best-fit power-law continuum ($\Gamma = 2.04^{+0.20}_{-0.15}$, \nH
  $=1.16^{+0.56}_{-0.36}$\psqcm). The line at around 6.5 keV is as
  strong as the 1993 ASCA line profile (Fig. 5). The EW is about 600
  eV.  Note also another line-like feature at 4.3 keV (see text for
  details).}
\end{figure}

The large EW of the Fe K$\alpha $ observed in the one-day ASCA
observation in 1993 may not be a very unusual phenomenon. The longer
ASCA observation in 1997 was examined for iron line variability. The
spectrum taken from the last 20 hr of the observation suggests that
the EW of the iron line could be as large as $615^{+814}_{-304}$ eV,
in contrast to the mean value $\sim 270$ eV (Table 2). During
this time interval, two flares were observed to show large amplitudes
(a factor of $\sim 2$, see Fig. 1 in Iwasawa et al 1998), but the mean
flux $1.83\times 10^{-11}$\ergpspsqcm\ is slightly below the average
of the whole observation (see Table 1).

A significant line-like feature is also detected at an energy of
$4.27^{+0.10}_{-0.10}$ keV. The energy is close to that of the ArXVIII
RRC, but the observed strength ($EW = 94^{+71}_{-48}$) is too large to
be identified with that feature. A possible alternative is a part of
gravitationally redshifted iron K$\alpha $ emission, if the line
emitting region is confined to a narrow ring at radii of $\sim 3$\Rs\ 
of a nearly face-on disk, which would produce a red horn at that
energy range. Fe K$\alpha $ emission propagating inward after a large
flash of X-ray illumination (reverberation) could produce such a
feature, although the narrowness of the feature (gaussian dispersion,
$\sigma = 0.17^{+0.19}_{-0.14}$) is not compatible with this
interpretation, as it should move across the energy band in a much
shorter time scale (1000 s or less, Young \& Reynolds 2000) for a
likely black hole mass of this object $\sim 10^7$\Ms. Another
interpretation is substructure in the line profile due to turbulence
in a magnetized disk (Armitage \& Reynolds 2003), which could account
for the 5.6 keV line observed in the spectrum of NGC3516 (Turner et al
2002).

We point out that a similar feature albeit at slightly higher
energy at $4.68^{+0.13}_{-0.10}$ keV is seen in the ASCA93
spectrum (see Fig. 5), which also exhibits a large Fe K$\alpha $ EW.
The 4.68 keV feature is unresolved (the 90 per cent upper limit for
gaussian dispersion is $\sigma = 0.22$ keV) and has a line flux of
$2.36^{+1.05}_{-1.23}\times 10^{-5}$\phpspsqcm\ ($EW\simeq 93$ eV).

The ionized reflection model (with the same parameter settings as the
fits shown in Table 3 with an addition of a gaussian for the 4.3 keV
line) gives $l_{\rm x}/l_{\rm d}=0.1 (\leq 0.4)$ with $\chi^2 = 742.7$
for 790 degrees of freedom. The small $l_{\rm x}/l_{\rm d}$ value is
in disagreement with that for the mean spectrum but similar to that of
the ASCA93 data, suggesting the accretion disk, as a reflecting
medium, might be in a similar state in the two occasions. The fact
that the averaged flux for the time interval during which the iron
line in Fig. 9 was observed differs very little from that of the rest
of the observation implies that the physical condition of the
reflector cannot be defined by an averaged source luminosity alone.
There might be some other factors which escapes from our observation.

\section{Discussion}

We have presented X-ray spectral evidence for highly photoionized matter
(Fe K$\alpha $ and SXVI RRC) in IRAS 18325--5926 and demonstrated that
its X-ray spectrum agree well with a model including reflection from
an ionized accretion disk. Relativistic blurring provides a good
explanation for the broad spectral features, particularly the iron
K$\alpha$ line, as expected at the inner radii of the accretion disk
around a central black hole. We have also tested for a gaussian smearing
of the ionized reflection features. The $\chi^2$ values obtained for
the spectral fits favours relativistic blurring over gaussian smearing
for all seven datasets. This indicates that the redward asymmetry,
characteristic to the relativistic effects, fits the data better than
the symmetric broadening.

A strong iron line from FeXXV as observed in IRAS 18325--5926 is not
very common among Seyfert galaxies. In the particular model for ionized
disks illuminated by an X-ray source by Nayakshin et al, the importance
of the thermal structure of the disks is emphasized for a strong ionized
iron line to be seen in the observed spectrum. The necessary conditions to
create a relatively cool layer of photoionized gas, which emits FeXXV
and FeXXVI, are met by steep X-ray continuum slope $\Gamma \simeq
2.2$ and a possible continuum roll-over at around 30 keV in IRAS
18325--5926.


We remains cautious about the reality of the continuum roll-over,
however. An immediate implication of the high-energy continuum
roll-over is that the temperature of the corona above the disk, which
Comptonizes soft photons, is cooler than that in other Seyfert
galaxies where the cut-off energy is $\sim 200 $ keV (e.g., Perola et
al 2002; Malizia et al 2003). For a photon index of $\Gamma = 2.2$ and
$kT_{\rm e} = 30$ keV, the optical depth of the corona to Thomson
(electron) scattering is $\tau_{\rm es}\simeq 1.7$ (e.g., Beloborodov
1999). If reflection from the disk passes through the corona before
reaching an observer, the spectrum of the reflection would be modified
greatly through Compton scatterings. This situation can be avoided by
the lamp-post type illumination geometry, where the illuminating
source is centrally concentrated above the disk. However, in
geometries with magnetized flares occurring in a corona just above the
disk, the spectral distortion via Compton scattering is inevitable.
For $kT_{\rm e}\sim 30 $ keV, the average fractional photon-energy
shift per scattering is $\sim 23$ per cent.  With $\tau_{\rm es} =
1.7$, a large fraction of photons from the disk would be scattered $N
= $ 2--3 times and double their energy (by a factor of $\exp [N
(4kT_{\rm e}/m_{\rm e}c^2)]$ where $N\sim\tau_{\rm es}^2$, Rybicki \&
Lightman 1979).  Expected effects on the observed spectrum are
broadened spectral features and reduction of the iron line EW by more
than half (e.g., Petrucci et al 2001). The Fe K$\alpha $ line profile
would be broadened significantly with $\sigma > 1$ keV and skewed to
higher energies (e.g., Pozdnyakov, Sobol \& Sunyaev 1979).

The fact that we sometimes observe a high EW of Fe K$\alpha $ as large
as 600 keV does not support that the reflection emission undergoes
significant Comptonization within the corona. If the EW of Fe K$\alpha
$ was controlled by the optical depth of the corona, then an
anti-correlation between line width and EW of Fe K$\alpha $ would be
expected, which is however not observed (see Table 2). The redward
asymmetry of the spectral features due to relativistic blurring, which
fits data well as mentioned above, is inconsistent with the expected
line profile emerging from the optically thick Comptonizing corona.
These would constrain the disk illumination geometry to a limited type
such as the lamp-post model (the patchy corona model may work as well,
since a fair fraction of the disk surface is left uncovered by the
corona), if the 30 keV roll-over is real. When a source is variable,
localized illumination of the disk expected in the patchy corona model
would lead to the reflected spectrum behaving differently from that in
the lamp-post model, e.g., a significant contribution from low
ionization reflection could be present, as discussed by Collin et al
(2003). Given the presence of relativisitic broadening, this is hard
to examine with the iron line emission. No significant cold reflection
is required by the high energy continuum observed with Ginga, RXTE and
BeppoSAX, although weak 6.4 keV line emission is seen in the high
quality XMM-Newton data.

The cool corona would mean that the production of high energy photons via
Compton up-scattering needs more scatterings, resulting in longer
delay of hard X-ray emission relative to softer X-ray emission, which
might be the case for the energy dependence of the X-ray light curves
obtained from the XMM-Newton observation (Iwasawa et al 2003 in
prep.).  Therefore given the implications mentioned above, verifying
the roll-over is highly desirable. It will probably have to wait until
the launch of ASTRO-E2.




Between the seven recent X-ray observations presented in this paper
spanning over 12 yr, the iron line flux has changed significantly in
response to the continuum change. The line flux appears to correlate
with the continuum flux in at least some cases, but the strength of
the line may be determined in a more complex way (e.g., Nayakshin \&
Kazanas 2002; Collin et al 2003), as the presence of occasional
outliers suggest (see also discussion in Section 3.8). The accretion
rate, disk illumination pattern, the physical conditions of the X-ray
emitting corona and the ionized surface of the disk, and also strong
relativistic effects (Cunningham 1976; Martocchia \& Matt 1996;
Miniutti et al 2003) etc.  may all play a role in controlling the
production of the iron line emission. As shown in Fig. 8, the
temperature of the disk surfaces inferred by $l_{\rm x}/L_{\rm d}$ may
be important. This parameter could be related to the fraction of
accretion power going into the corona, which changes in time. Further
complexity is added by the time dependence of some of those properties
mentioned above. Since rapid variability is common in this source
(see, e.g., Fig. 1), the X-ray source is unlikely to be in a steady
state.  In this paper, we have investigated data averaged over the
durations of half to a few days, for which the time evolution in
heating and cooling of the Comptonizing corona (e.g., Guilbert, Fabian
\& Ross 1982) should be averaged over each observed duration, whilst
the dynamical time scale of the accretion flow is perhaps relevant.
The unusually large EW of the line appears only in relatively short
time intervals ($\leq 1$ day) as discussed in Section 3.8, which may
be intriguing in this respect.

In the presence of strong reflection from a highly ionized disk,
nearly half of the observed X-ray flux, e.g., in the 2--10 keV band, is
attributed to the reflection. This would reduce the estimate of the
intrinsic luminosity of the illuminating X-ray source by approximately
half.

\section*{Acknowledgements}

Sergei Nayakshin and Matteo Guainazzi are thanked for useful discussion. ACF
and KI acknowledge Royal Society and PPARC, respectively, for support.
AJY and CSR acknowledge support from NASA grant NAG5-9935 and the
National Science Foundation grant AST0205990, respectively.  JCL
acknowledges support from the Chandra Fellowship grant PF2-30023 --
this is issued by the Chandra X-ray Observatory Center, which is
operated by SAO for and on behalf of NASA under contract NAS8-39073.

\end{document}